\documentclass[final,5p,xchauthor,chkrefs,GCNS,amsmath,times,fleqn]{elsarticle}

\usepackage{graphicx}
\usepackage{amssymb}
\usepackage{mathrsfs}
\usepackage{amsmath}
\usepackage{comment}
\usepackage{amsbsy}
\usepackage{amsfonts}
\usepackage{lineno}
\usepackage{nicefrac}
\usepackage{epsf,color,colordvi,pifont}
\biboptions{sort&compress}
\usepackage[notref,notcite]{showkeys} 
\def\deltabar{{\mathchar'26\mkern-7.5mu\delta}}
\def\dbar{{\mathchar'26\mkern-11.8mu d}}
\def\lambdabar{{\mkern0.75mu\mathchar'26\mkern-9.75mu\lambda}}
\DeclareFontFamily{OT1}{pzc}{}
\DeclareFontShape{OT1}{pzc}{m}{it}{<-> s * [1.10] pzcmi7t}{}
\DeclareMathAlphabet{\mathpzc}{OT1}{pzc}{m}{it}

\def\dbar{{\mathchar'26\mkern-12mu d}}
\begin{document}

\begin{frontmatter}
\title{Minicharged particles search by strong laser pulse-induced vacuum polarization effects}
\author[UHH]{S. Villalba-Ch\'avez}
\ead{selym@tp1.uni-duesseldorf.de}
\author[MPI]{S. Meuren}
\ead{sebastian.meuren@mpi-hd.mpg.de}
\author[UHH]{C. M\"{u}ller}
\ead{c.mueller@tp1.uni-duesseldorf.de}
\address[UHH]{Institut f\"{u}r Theoretische Physik I, Heinrich-Heine-Universit\"{a}t D\"{u}sseldorf, Universit\"{a}tsstr. 1, 40225 D\"{u}sseldorf, Germany}
\address[MPI]{Max-Planck-Institut f\"{u}r Kernhysik, Saupfercheckweg 1, 69117 Heidelberg, Germany}
\date{\today}

\begin{abstract}
Laser-based searches  of   the  yet unobserved vacuum birefringence might be sensitive for very light hypothetical  particles carrying a tiny fraction of the electron charge.
We show that,  with the help of contemporary techniques, polarimetric  investigations  driven by an optical laser pulse of moderate intensity  might allow for  excluding 
regions of the parameter space of these particle candidates  which have  not been discarded so far by laboratory measurement data. Particular attention is paid to the role of a  
Gaussian wave profile. It is argued that, at  energy regimes in which the vacuum becomes dichroic due to these minicharges,  the transmission probability of a probe beam  through an analyzer set 
crossed to the  initial polarization direction  will depend on both the induced ellipticity as well as   the rotation of  the initial   polarization plane. The weak 
and strong field regimes,  relative to the attributes of these minicharged particles,  and   the relevance of the  polarization of the strong field are  investigated. 
\end{abstract}
\end{frontmatter}

\section{Introduction}

The Standard Model  of particle physics is currently understood as an effective theory, where  charge quantization seems to be  
conceived as a fundamental principle. Standard Model extensions--which are required  for other reasons--can be found either by  
enforcing the  mentioned quantization through higher gau\-ge groups or  by incorporating  carriers of a tiny  charge $\mathpzc{q}_\epsilon=\epsilon \vert e\vert$, 
with  $\epsilon$ denoting  the  parameter relative to the absolute value of the   electron charge   $e<0$  \cite{Batell:2005wa,Bruemmer:2009ky,Brummer:2009cs,Dudas:2012pb,Goodsell:2009xc,Langacker:2008yv}.  
That these particle  candidates have eluded a  direct experimental verification  indicates that their interaction with the well established 
Standard Model  branch might be extremely feeble [$\epsilon\ll1$]. In light of this situation, the parameter space of this sort of Mini-Charged 
Particles (MCPs) \cite{Okun:1982xi,Holdom:1985ag,Dobrescu:2006au,Gies:2006ca,Ahlers:2007rd,Ahlers:2007qf} is  being limi\-ted. Stringent constraints have been inferred from nonobservable 
effects in  the stellar evolution   \cite{davidson} [$\epsilon\lesssim 10^{-14}$ for masses $m_\epsilon$ below a few $\rm keV$]  and the analysis of   
the big bang nucleo\-synthesis [$\epsilon<10^{-9}$ for $m_\epsilon< 1\ \rm MeV$]. However, these  astro-cosmological bounds are  somewhat 
vulnerable due to the uncertainty  associated with the underlying phenomenological model  \cite{Masso:2005ym,Masso:2006gc,Jaeckel:2006id,evading}.  
Laboratory limits are considerably less stringent but  more reliable.  They have been  established from  
regeneration  setups  \cite{Ehret:2010mh,Chou:2007zzc,Steffen:2009sc,Afanasev:2008jt,Pugnat:2007nu,Robilliard:2007bq,Fouche:2008jk},\footnote{An alternative regeneration setup based 
on  static magnetic fields has been proposed in Ref.~\cite{Dobrich:2012sw}.} tests for mo\-difications in  
Coulomb's law \cite{Jaeckel:2009dh,Jaeckel:2010xx}  or through high precision experiments  looking for magnetically-induced vacuum   birefringence and 
vacuum  dichroism \cite{Cameron:1993mr,PVLAS2008,DellaValle:2013xs,BMVreport,Chen:2006cd}.\footnote{A more extended phenomenological overview on 
MCPs as well as other weakly interacting particles   can be found in the  reviews  \cite{Jaeckel:2010ni,Ringwald:2012hr,Hewett:2012ns,Essig:2013lka}.} 
In the last scenario  the bound is the more stringent  the greater  the  field strength and  its spatial extension are.  However, in laboratories, the highest constant 
magnetic fields do  not exceed  values  of  the order of  $\sim10^{6}\ \mathrm{G}$, which are extended over effective distances  of upto  $10-100$ kilometers 
using  Fabry-P\'erot cavities.

Fields generated from high-intensity lasers might be beneficial for these laboratory searches.  Indeed,  the chirped-pulse amplification technique  
has enabled us to reach very strong magnetic field strengths, at the expense  of being distributed inhomogeneously over regions of only a few micrometers 
\cite{Di_Piazza_2012}. Strengths as large as $\sim 10^9\ \rm G$ are accessible  nowadays and will likely exceed 
values of the order of  $\sim 10^{11}\ \rm G$  at forthcoming  laser systems  such as ELI and XCELS \cite{ELI,xcels}. This fact also justifies why  high-intensity 
laser pulses are currently  considered as valuable  instruments for detecting various nonlinear phenomena that have eluded  
their observation so far. Notably, to measure vacuum birefringence \cite{Heinzl,DiPiazza:2006pr,Dinu:2013gaa,Dinu:2014tsa,King:20161},   the 
HIBEF consortium has  proposed a laser-based polarimetric  experiment which combines a Petawatt optical  laser with a x-ray free electron laser \cite{HIBEF,Schlenvoigt}. 
Meanwhile,  alternative setups are being proposed for improving   the levels of sensitivity  necessary  for  the  detection of  this elusive   phenomenon  \cite{Tommasini:2009nh,Karbstein:2015xra,King:2016jnl,Nakamiya:2015pde}. 
Clearly, experiments of this nature might also constitute sensitive probes for  axion-like particles \cite{mendonza,Gies:2008wv,Dobrich:2010hi,Villalba-Chavez:2013bda,Villalba-Chavez:2013goa}, 
MCPs and  paraphotons \cite{Villalba-Chavez:2013gma,Villalba-Chavez:2014nya,Villalba-Chavez:2015xna,Gabrielli:2016rhy}. This forms  the main motivation for this work. 
In this Letter we show that a  polarimetric probe  driven by the field of a high intensity linearly polarized Gaussian laser pulse might  notably  improve the 
existing  laboratory  limits in some regions of the parameter space of MCPs.

Our investigation relies on the one-loop  representation of  the polarization  tensor in a plane-wave background  \cite{baier,Mitter,Meuren:2013oya}  in which the two-point correlation  
function  for MCPs incorporates the field of the laser  pulse  in a nonpertubative way  [Furry picture]. The weak and strong field regimes, relative to the 
attributes of these degrees of freedom, are investigated and asymptotic expressions  for  the observables are derived [see Sec.~\ref{secasymp} for more details]. In the weak field 
case,  dispersive effects are found to be  maximized at the threshold of pair production of MCPs,  in agreement with the cross section of  light-by-light scattering. Finally, a 
comparison between the present results and those  previously obtained for a circularly polarized   monochromatic plane-wave background \cite{Villalba-Chavez:2014nya,Villalba-Chavez:2015xna} is established.

\section{Photon propagation in MCPs vacuum \label{GP}}

We wish  to evaluate the effects induced by  quantum vacuum fluctuations dominated by  Dirac fields  characterized by a mass $m_\epsilon$ and  a tiny fraction of the absolute value of the electron charge
$\mathpzc{q}_\epsilon\equiv\epsilon\vert e\vert$. As long as such fields are minimally coupled to an  electromagnetic field and the corresponding functional action preserves the formal 
invariance properties of Quantum Electrodynamics (QED),  the underlying theory would resemble the corresponding  phenomenology. Accordingly, the  equation of motion--up to linear terms in the small-amplitude 
wave  $a_\mu(x)$--has  the form\footnote{From now on ``natural'' and Gaussian  units  $c=\hbar=4\pi\epsilon_0=1$ are used.}
\begin{eqnarray}
\square a_\mu(x)+\int d^4x^\prime\Pi_{\mu\nu}(x,x^\prime)a^{\nu}(x^\prime)=0,  \label{DSE}
\end{eqnarray}  provided  the Lorenz gauge $\partial_\mu a^\mu =0$ is chosen. Here, $\square\equiv\partial_\mu\partial^\mu=\partial^2/\partial t^2-\nabla^2$, whereas   
the second term in Eq.~(\ref{DSE}) introduces  the vacuum polarization tensor  $\Pi_{\mu\nu}(x,x^\prime)$. This object is basically the same  as in  QED, with the  positron  parameters $(\vert e\vert,\ m)$  
substituted by the respective quantities associated with the MCP $(\mathpzc{q}_\epsilon,\ m_\epsilon)$. It constitutes the lowest nontrivial one-particle irreducible vertex from which 
the gauge sector of QED can  acquire a dependence on the external  background field. Its four-potential  is taken hereafter as
\begin{equation}\label{extfield}
\mathscr{A}^{\mu}(x)=\mathpzc{a}^{\mu}_{1}\psi_1(\varphi)+\mathpzc{a}^{\mu}_{2}\psi_2(\varphi),
\end{equation} where $\mathpzc{a}_{1,2}$ are two orthogonal amplitude vectors [$\mathpzc{a}_{1}\mathpzc{a}_{2}=0$]  and   $\psi_{1,2}(\varphi)$  arbitrary  functions of the  strong 
plane-wave phase $\varphi=\varkappa x$.  The external potential is chosen  in  the Lorenz  gauge $\partial_{\mu}  \mathscr{A}^\mu=0$  so that   the wave four-vector  $\varkappa^{\mu}=(\varkappa^0,\pmb{\varkappa})$ 
 with $\varkappa^2=0$  and  the   amplitude  vectors  $\mathpzc{a}_{1,2}^\mu$   satisfy the constraints  $\varkappa \mathpzc{a}_{1,2}=0$. 

At this point, it turns out to be rather useful to introduce the four-vectors \cite{baier}
\begin{eqnarray}\label{vectorialbasisbaier}
\Lambda_{1,2}^\mu(q)=-\frac{\mathscr{F}_{1,2}^{\mu\nu}q_\nu}{\varkappa q\sqrt{-\mathpzc{a}_{1,2}^2}},\quad \Lambda_{3,4}^\mu(q_{1,2})=\frac{\varkappa_\mu q_{1,2}^2-q_{1,2}^{\mu}(q\varkappa)}{\varkappa q\sqrt{q_{1,2}^2}},
\end{eqnarray} which are built up from  the amplitudes of the  external  field modes  $\mathscr{F}^{\mu\nu}_{i}=\varkappa^\mu\mathpzc{a}^\nu_{i}-\varkappa^\nu\mathpzc{a}^\mu_{i}$ [$i=1,2$], the respective 
incoming and outgoing four-momenta of the probe photons $q_1$ and $q_2$ as well as the wave four-vector $\varkappa$. We note that the shorthand notation $q$ in Eq.~\eqref{vectorialbasisbaier} may stand 
for either $q_1$ or $q_2$ due to momentum conservation. The set of  four-vectors $q_1$,  $\Lambda_{1}(q_1)$, $\Lambda_{2}(q_1)$ and $\Lambda_{3}(q_1)$,   form a complete orthonormalized basis, 
i.e., $\Lambda_{i}^{\mu}(q_1)\Lambda_{j \mu}(q_1)=-\delta_{ij}$,  $\mathpzc{g}^{\mu\nu}=q_1^{\mu} q_1^{\nu}/q_1^2-\sum_{i=1}^{3}\Lambda_i^{\mu }(q_1)\Lambda_i^{\nu }(q_1)$ with $\mathpzc{g}_{\mu\nu}=\mathrm{diag}(+1,-1,-1,-1)$ 
denoting the metric tensor. A similar statement applies to the set of four-vectors  $q_2$,  $\Lambda_{1}(q_2),$ $\Lambda_{2}(q_2)$ and $\Lambda_{4}(q_2)$. 

Let us proceed by Fourier transforming Eq.~(\ref{DSE}). In the following we will seek the solutions  of the resulting equation in the form of a superposition of transverse waves  
$a^\mu(q)=\sum_{i=1,2}\Lambda^\mu_{i}(q)f_i(q)$. Correspondingly, 
\begin{equation}
\begin{split}
&q_2^2f_i(q_2)=-\sum_{j=1,2}\int \dbar^4 q_1\Lambda_{i}^\mu(q_2)\Pi_{\mu\nu}(-q_2,-q_1)\Lambda^\nu_j(q_1)f_j(q_1),\\
&\Pi_{\mu\nu}(q_1,q_2)=\frac{\deltabar_{q_2,q_1}}{\varkappa_+}\int d\varphi \mathpzc{P}_{\mu\nu}(\varphi,q_1, q_2)\exp\left[\frac{i(q_2-q_1)_+}{\varkappa_+}\varphi\right],
\end{split} \label{fouriertransf}
\end{equation}where we have introduced the shorthand notations   $\quad \dbar\equiv d/(2\pi)$ and $\quad \deltabar_{q_2,q_1}\equiv(2\pi)^3\delta^{(\perp)}(q_2-q_1)\delta^{(-)}(q_2-q_1)$. 
Note that  quantities with subindices   $\pm$ and $\perp$   refer  to  light-cone coordinates. We  choose  the  reference frame in such a way  that the direction of  propagation of our  
external plane wave  [see Eq.~(\ref{extfield})] is along the positive direction of the  third axis. As  a consequence,   the strong field only depends on  $x_-=(x^0-x^3)/\sqrt{2}$ via $\varphi=\varkappa_+x_-$  
with  $\varkappa_+=(\varkappa^0+\varkappa^3)/\sqrt{2}=\sqrt{2}\varkappa_0>0$ and  the remaining light-cone variables, i.e.  $x_+=(x^0+x^3)/\sqrt{2}$ and $\pmb{x}_\perp=(x^1,x^2)$ 
can be integrated out without  complications. 

Although the  expression above holds for arbitrary external field profiles,  it still requires a transversely homogeneous field. As a consequence,  $\pmb{q}_{\perp}$ is  conserved 
[see the associated Dirac delta in Eq.~(\ref{fouriertransf})], which  constitutes  a good approximation whenever the Compton wavelength of the MCP  $\lambdabar_\epsilon=1/m_{\epsilon}$  becomes 
much smaller than the transverse  length scale  over which the field is  homogeneous. For a focused laser beam this scale is set by the waist size   of  the pulse $w_0$. Therefore, the plane-wave 
approximation is valid in the regime  $m_\epsilon \gg w_0^{-1}$.  The study of the  regime $m_\epsilon \lesssim w_0^{-1}$,  where spatial focusing effects 
become important, is beyond the scope of the present investigation. 

The  tensorial structure of $\mathpzc{P}_{\mu\nu}(\varphi,q_1, q_2)$ can be determined on the basis of symmetry principles, independently of any approximation 
used to compute the polarization tensor \cite{baier,Meuren:2013oya}. It  reads
\begin{equation}
\begin{split}
&\mathpzc{P}^{\mu\nu}(\varphi,q_1, q_2)=c_1\Lambda^\mu_{1}\Lambda^\nu_{2}+c_2\Lambda^\mu_{2}\Lambda^\nu_{1}+c_3\Lambda^\mu_{1}\Lambda^\nu_{1}\\&\qquad\qquad\qquad\qquad\qquad\qquad\quad+c_4\Lambda^\mu_{2}\Lambda^\nu_{2}+c_5\Lambda^\mu_{3}\Lambda^\nu_{4}.
\end{split}\label{DVPBaier}
\end{equation} As  $q_1-q_2\sim \varkappa$ this   decomposition  does not depend on which choice of $q$ is taken; see also Eq.~(\ref{vectorialbasisbaier}). The form factors  $c_{i}$    in Eq.~\eqref{DVPBaier} 
depend--among other parameters--on the phase of the external field $\varphi$,   $q_1$ and $q_2$. 
In the one-loop approximation--which  is adopted from now on--they turn out to be  represented  by two-fold parametric integrals in the  variables  $\tau\in[0,\infty)$ and $v\in[0,1]$,  the integrand 
of which being of the form [see Ref.~\cite{baier}]
\begin{equation}\label{integrandform}
\exp[-im_\epsilon^2\tau+i\mu q_1^2]\times \left(\mathrm{Regular}\ \mathrm{Function}\ \mathrm{in}\ q_1^2,\ q_2^2\ \mathrm{and}\ \varkappa q\right)
\end{equation}
with  $\mu=\frac{1}{4}\tau(1-v^2)$. After a suitable integration by parts  the regular function becomes  independent of $q_2^2$  [see Ref.~\cite{Meuren:2014uia}, App.~D for more details],  which is assumed in the following.  
  
When  polarization effects  do not dramatically modify the photon dispersion law in vacuum  [$q^2=0$],  one can  solve Eq.~(\ref{fouriertransf}) perturbatively by setting 
$f_{i}(q)\approx f_{0i}(q)+\delta f_i(q)$. In the following, we suppose a head-on  collision between  the strong laser pulse and the probe beam characterized by the four-momentum  
$k^\mu=(\omega_{\pmb{k}},\pmb{k})$,  so that $\varkappa_+k_{-}=2\omega_{\pmb{k}}\varkappa_0$ and $\pmb{k}_{\perp}=\pmb{0}$. 
Accordingly, the leading order term is $f_{0i}(q)=\vert 2q_-\vert a_{0 i}\ \deltabar(q^2)\ \deltabar^{(\perp)}(q)\ \deltabar^{(-)}(q-k)$, corresponding to  $f_{0i}(x)=a_{0i}e^{-i\phi}$ 
with  $\phi=kx=k_{-}x_+$ and $a_{0i}$ the  amplitude of mode-$i$. Then, it follows from Eq.~(\ref{fouriertransf}) that  the perturbative contribution is given by
\begin{equation}
\begin{split}
&\delta f_i(q_{2})=-[2q_{2+}q_{2-}-q_{2\perp}^2+i0]^{-1}\\&\qquad\qquad\qquad\qquad\times\sum_{j=1,2}a_{0j}\Lambda_{j}^\mu(k)\Pi_{\mu\nu}(k,q_2) \Lambda^\nu_i(q_2),
\end{split}
\end{equation}where it must be understood that the only nonvanishing light-cone component of the four-vector $k^\mu$   is $k_-$.  
Besides, in obtaining the expression above we have used the symmetry property $\Pi_{\mu\nu}(-q_2,-q_1)=\Pi_{\nu\mu}(q_1,q_2)$. 
Here, the poles in the function  $1/q_2^2$ have been shifted  infinitesimally into the complex plane  by an $i0$-term  so that   correct boundary conditions  of the fields
at asymptotic times  $f_i(\pm\infty,\pmb{x})$ are implemented.  In this case,   the solution of Eq.~(\ref{DSE})  is given by $a^\mu(x)=\sum_{i=1,2}\Lambda_i^\mu(k) f_i(x)$  [see above Eq.~(\ref{fouriertransf})]  with
\begin{equation}\label{intermediateequation}
\begin{split}
&f_i(x)\approx f_{0i}(x)-\frac{1}{2\varkappa_{+}k_{-}}\sum_{j=1,2}f_{0j}(x)\int d\tilde{\varphi}\int \dbar q_{2+} \\&\quad\qquad\qquad\times e^{\frac{iq_{2+}}{\varkappa_+}(\tilde{\varphi}-\varphi)} \Lambda_{j}^\mu(k)\frac{\mathpzc{P}_{\mu\nu}(\tilde{\varphi},k,q_{2})}{q_{2+}+i0} \Lambda^\nu_i(q_2).
\end{split}
\end{equation}Here, $q_{2-}=k_-$, $\pmb{q}_{2\perp}=\pmb{0}$,  whereas  $\pmb{k}_\perp=\pmb{0}$ and $k_+=0$. In order to provide  a more concise  expression for  $f_i(x)$,   we integrate out  $q_{2+}$. 
This can be carried out by applying  Cauchy's theorem  and  the residue theorem, depending upon whether the contour of integration is  chosen  in  the upper or lower  half of the complex plane. Taking into account the 
structure  of the  integrand  with respect to $q_{2+}$ [see Eq.~(\ref{integrandform}) and the discussion below], we obtain
\begin{equation}
 \int \dbar q_{2+}\ldots=-i\Lambda_{j}^\mu(k)\mathpzc{P}_{\mu\nu}(\tilde{\varphi},k,k) \Lambda^\nu_i(k)\Theta(\varphi-\tilde{\varphi}),
\end{equation}where  $\Theta(x)$ denotes the unit step function. Its emergence  restricts the integral over  $\tilde{\varphi}$ to    $(-\infty,\varphi]$ instead of $(-\infty,\infty)$, 
as required by causality. However, we are only interested in  asymptotically  large spacetime distances  [$\varphi\to\infty$], i.e.,  when the  high-intensity  laser  field  is turned off,  
which  restores  the original integration limits.   Therefore, inserting this expression into Eq.~(\ref{intermediateequation}) and taking into account the tensorial decomposition of 
the polarization tensor [see Eq.~(\ref{DVPBaier})], 
we end up with
\begin{equation}\label{degradedDSE}
\begin{split}
&f_i(x)\approx f_{0i}(x)+\frac{i}{2\varkappa_{+}k_{-}}f_{01}(x)\int_{-\infty}^{\varphi} d\tilde{\varphi}\left[c_3(\tilde{\varphi})\delta_{i1}+c_1(\tilde{\varphi})\delta_{i2}\right]
 \\&\qquad\qquad+\frac{i}{2\varkappa_{+}k_{-}}f_{02}(x)\int_{-\infty}^{\varphi} d\tilde{\varphi}\left[c_4(\tilde{\varphi})\delta_{i2}+c_2(\tilde{\varphi})\delta_{i1}\right].
\end{split}
\end{equation} The expression above  constitutes the starting point for further considerations. It holds for arbitrary strength and polarization of the background  field, as long as the vacuum polarization is small.  
When specifying  Eq.~(\ref{degradedDSE}) to the case of a  linearly polarized  plane-wave background, i.e. Eq.~(\ref{extfield}) with $\psi_2(\varphi)=0$, the form factors $c_{1,2}$ vanish   
\cite{baier,Meuren:2013oya} and the resulting expression agrees with  Eq.~(16) in Ref.~\cite{Meuren:2014uia}, provided the involved exponential function  is expanded to leading order. However, we 
emphasize that the  aforementioned solution  has been established for the field regime in which the    laser  intensity parameter $\xi=\vert e\vert\sqrt{-\mathpzc{a}^2}/m$ with $\mathpzc{a}^\mu\equiv \mathpzc{a}_1^\mu$ 
is  very large [$\xi\gg1$]. 

Now, if the external field is linearly polarized, the solution of Eq.~(\ref{degradedDSE}) allows us to  write the  electric field of the probe  [$\pmb{\varepsilon}(x)=-\partial\pmb{a}/\partial x^0$ with $a_0=0$] 
as a superposition of plane-waves
\begin{eqnarray}
\begin{split}
&\pmb{\varepsilon}(x)\approx\varepsilon_0\cos(\vartheta_0)\pmb{\Lambda}_{1} \mathrm{Re}\, e^{-i\phi+\frac{i}{2\varkappa_+ k_-}\int_{-\infty}^{\varphi} d\tilde{\varphi}\ c_3(\tilde{\varphi})}\\&\qquad\qquad +
\varepsilon_0\sin(\vartheta_0)\pmb{\Lambda}_{2}\mathrm{Re}\, e^{-i\phi+\frac{i}{2\varkappa_+ k_-}\int_{-\infty}^{\varphi} d\tilde{\varphi}\ c_4(\tilde{\varphi})}. 
\end{split}\label{electricfield}
\end{eqnarray}Here,  $\varepsilon_0$ refers to the initial  electric field amplitude, $\pmb{\Lambda}_{1,2}=\pmb{\mathpzc{a}}_{1,2}/\vert\pmb{\mathpzc{a}}_{1,2}\vert$,  whereas  $0\leqslant\vartheta_0<\pi$ is 
the  corresponding  initial polarization  angle  of the probe with  respect to $\pmb{\Lambda}_{1}$, i.e., the polarization axis of the external pulse. Observe  that the 
appearance of the phase is due to the approximation $1 + ix \approx \exp(ix)$  as in Ref.~\cite{Dinu:2013gaa}. 

The  $\mathpzc{P}_{\mu\nu}-$form factors  are, in  general,  complex  functions $c_{3,4}=\mathrm{Re}\ c_{3,4}+i\ \mathrm{Im}\ c_{3,4}$. Correspondingly, the exponents in 
Eq.~(\ref{electricfield}) contain  real and imaginary contributions. The latter are connected to the photo-production of MCP pairs via the optical theorem \cite{Meuren:2014uia,VillalbaChavez:2012bb}, 
a phenomenon  which damps  the intensity of the probe,  $I(\varphi)=\frac{\varepsilon_0^2}{4\pi} \cos^2(\vartheta_0) \exp(-\kappa_{1})+\frac{\varepsilon_0^2}{4\pi} \sin^2(\vartheta_0) \exp(-\kappa_{2})$, as it propagates in the pulse.  
As such, the analytic properties of the  factors $\kappa_{1,2}\equiv\kappa_{1,2}(\varphi)=\frac{1}{\varkappa_+ k_-}\mathrm{Im}\int_{-\infty}^{\varphi} d\tilde{\varphi}\ c_{3,4}(\tilde{\varphi})$, responsible for the damping   
differ from each other, leading to a nontrivial difference $\delta\kappa(\varphi)=\frac{1}{\varkappa_+k_-}\mathrm{Im}\ \Delta(\varphi)$, where we introduced  the function
\begin{equation}
\Delta(\varphi)=\int_{-\infty}^{\varphi} d\tilde{\varphi}\ \left[c_{3}(\tilde{\varphi})-c_4(\tilde{\varphi})\right].\label{difference}
\end{equation} Therefore,  the vacuum  behaves  like  a dichroic  medium, inducing a rotation of the probe polarization  from the initial angle  $\vartheta_0$ to $\vartheta_0+\delta\vartheta$, where  
$\delta \vartheta$ is expected to be tiny. At asymptotically  large spacetime distances  [$\varphi\to\infty$],  we find 
\begin{eqnarray}\label{rot}
\vert\delta\vartheta(\epsilon,m_\epsilon)\vert\approx \frac{1}{2}\sin(2\vartheta_0)\left\vert\frac{\mathrm{Im}\ \Delta(\infty)}{2\varkappa_+k_-}\right\vert\ll1.
\end{eqnarray} As the phase difference  between the two propagating modes, 
$\delta\phi(\varphi)=\frac{1}{2\varkappa_+ k_-}\mathrm{Re}\ \Delta(\varphi)$,  does not vanish  either [see Eq.~(\ref{electricfield})],  the vacuum is also predicted  to be  birefringent. 
Hence,  when the strong field is turned off [$\varphi\to\infty$],  
the  outgoing probe should be elliptically polarized and its  ellipticity   is given by \cite{Born} [note that in this reference a  different notation is used]
\begin{equation}\label{elip}
\vert\psi(\epsilon,m_\epsilon)\vert\approx \frac{1}{2}\sin(2\vartheta_0)\left\vert\frac{\mathrm{Re}\ \Delta(\infty)}{2\varkappa_+k_-}\right\vert\ll1.
\end{equation} In the case of optical probes, isolated  detections of  the rotation effect [see Eq.~(\ref{rot})] and the ellipticity [see Eq.~(\ref{elip})] could  be carried out depending 
on  whether a quarter wave  plate  is inserted  or not  in the path of the outgoing probe beam in front of a Faraday cell and an analyzer \cite{Cameron:1993mr,PVLAS2008,Chen:2006cd}.   
The latter is set crossed to the initial direction of polarization so that the transmitted photons are polarized  orthogonally. Correspondingly, no photons are detected in the
absence of birefringence and dichroism. Using high-purity polarimetric techniques for x-rays  \cite{Marx2011,Marx2013} (QED) vacuum birefringence could also be measured with a similar setup
by combining a x-ray probe and a strong optical field [QED-induced dichroism is exponentially small, thus $\delta\vartheta_{\mathrm{QED}}=0$ for practical purposes].
Such an experiment is envisaged at HIBEF \cite{Schlenvoigt}. 

In a scenario including  MCPs, the analysis must be revisited.  To this end, let us  consider the scattering amplitude   
$T=i \mathpzc{e}_\mu^{(i)}\left[\Pi_{\mathrm{QED}}^{\mu\nu}(k_1,k_2)+\Pi^{\mu\nu}(k_1,k_2)\right]\mathpzc{e}_\nu^{(f)}/[2V(\omega_{\pmb{k}_1}\omega_{\pmb{k}_2})^{\nicefrac{1}{2}}]$. The expression above includes both, the polarization tensor associated with QED $\Pi_{\mathrm{QED}}^{\mu\nu}(k_1,k_2)$ and the one related to the MCPs. Besides, 
$V$ denotes  the normalization volume, whereas  $\mathpzc{e}^{(i)}_\mu$ and $\mathpzc{e}^{(f)}_\mu$ are  the initial  and final polarization states,  respectively. Following 
Eq.~(\ref{electricfield}), we  suppose that  the former  is of the form   $\mathpzc{e}^{(i)}=\cos(\vartheta_0)\Lambda_1+\sin(\vartheta_0)\Lambda_2$. In contrast, the  polarization state transmitted by the analyzer is   
$\mathpzc{e}^{(f)}=\pm\sin(\vartheta_0)\Lambda_1\mp\cos(\vartheta_0)\Lambda_2$,  so that $\mathpzc{e}^{(i)}\mathpzc{e}^{(f)}=0$.  Finally, we establish the 
following expression for the transmission probability 
[$\delta\vartheta_{\mathrm{QED}}=0$]:
\begin{equation}\label{flippingprobability}
\begin{split}
&\mathcal{P}
=\left[\psi_{\mathrm{QED}}+\psi(\epsilon,m_\epsilon)\right]^2+\delta\vartheta(\epsilon,m_\epsilon)^2. 
\end{split}
\end{equation}This expression  indicates that the described  setup is not suitable to probe the signals separately. However, one could  achieve this goal by  determining the local minimum of the 
count rate behind the analyzer,  which is no longer perpendicular to the incoming polarization direction but shifted  by $\delta\vartheta(\epsilon,m_\epsilon)$  \cite{Schlenvoigtc}. We indeed find that  
in such a configuration,  the transmission probability  $\mathcal{P}_{\mathrm{min}}=\vert \pmb{\mathpzc{e}}\cdot \pmb{\varepsilon}\vert^2/\vert \varepsilon_{0}\vert^2$ with $\pmb{\mathpzc{e}}=\pm\sin(\vartheta_0+\delta\vartheta)\pmb{\Lambda}_1\mp\cos(\vartheta_0+\delta\vartheta)\pmb{\Lambda}_2$, 
is given by  the first term on the right-hand side of Eq.~(\ref{flippingprobability}). In connection, the number of photons transmitted  through the analyzer reads $
\mathcal{N}\approx\mathcal{N}_{\mathrm{in}}\mathcal{N}_{\mathrm{shot}}\mathcal{T}\left[\psi_{\mathrm{QED}}^2+2\psi_{\mathrm{QED}}\psi(\epsilon,m_\epsilon)\right]$, provided that QED 
effects are  dominant [$\psi_{\mathrm{QED}}>\psi(\epsilon,m_\epsilon)$]. Here, $\mathcal{N}_{\mathrm{shot}}$  counts the number of laser shots used for a measurement,  $\mathcal{T}$ 
denotes  the transmission coefficient of  all optical components  and $\mathcal{N}_{\rm in}$ is the  number of incoming  x-ray probe photons, respectively.  

\section{Asymptotic regimes \label{secasymp}}

We  wish  to investigate  the  optical  observables [Eq.~(\ref{rot}) and (\ref{elip})] induced by a plausible existence of  MCPs. Since  
both  depend on $\Delta(\infty)$  [see Eq.~(\ref{difference})],   we will focus on determining  this function. Indeed, a suitable expression 
can  be  inferred  from the literature \cite{baier,Meuren:2013oya}. In the one-loop approximation we find:
\begin{eqnarray}
\begin{split}
&\Delta(\infty)=\frac{\alpha_\epsilon}{\pi}m_\epsilon^2 \xi_{\epsilon}^2\int_{-\infty}^\infty d\varphi\int_{-1}^1dv\\ &\qquad\qquad\qquad\times\int_0^\infty \frac{d\tau}{\tau} X(\varphi) \exp\left[-im_{*}^2(\varphi)\tau\right],
\end{split}
\label{differneceformfactors}
\end{eqnarray}where  $\alpha_\epsilon\equiv \epsilon^2 e^2\approx\epsilon^2/137$ denotes   the fine structure constant relative to the MCPs, whereas  $\xi_{\epsilon}=\epsilon m \xi/m_\epsilon$ 
is  the relative  intensity parameter.  The remaining  functions  involved in this expression can be conveniently written in the following form
\begin{eqnarray}
\begin{split}
&X(\varphi)=\mu^2(2\varkappa_+k_-)^2\int_0^1 dy \int_0^1 d\tilde{y}\ y (\tilde{y}-1)\psi^\prime(\varphi_y)\psi^\prime(\varphi_{\tilde{y}})  ,\\ 
&m_*^2(\varphi)=m_\epsilon^2\left\{1-\xi_\epsilon^2\mu^2(2 \varkappa_+k_-)^2\int_{0}^{1} dy\ y\psi^\prime(\varphi_y)\right.\\&\qquad\qquad\qquad\times\left.\left[\int_0^1d\tilde{y}\ \tilde{y}\psi^\prime(\varphi_{\tilde{y}})-2\int_y^1d\tilde{y}\ \psi^\prime(\varphi_{\tilde{y}})\right]\right\}, 
\end{split}\label{parameters}
\end{eqnarray} where  $\mu=\frac{1}{4}\tau(1-v^2)$ and $\varphi_y=\varphi-2(\varkappa_+k_-)\mu y$. These expressions apply for a linearly  polarized plane-wave background [$\psi_1(\varphi)\equiv\psi(\varphi)$ 
and $\psi_2(\varphi)=0$]. Here, the prime denotes the derivative with respect  to the argument. An exact  evaluation of  $\Delta (\infty)$ [see Eq.~(\ref{difference})] is quite difficult to perform.  Therefore, 
we consider now  some asymptotic expressions of interest. 

\subsection{Leading behavior at  large $\xi_\epsilon\gg1$}

In order to  elucidate the asymptotic  contribution  of Eq.~(\ref{differneceformfactors}) at asymptotically large $\xi_\epsilon\gg1$  we first perform the  change 
of variable $\tau=4\rho/[\vert\varkappa_+k_-\vert(1-v^2)]$. The resulting integration over  $\rho$  is  divided into two contributions whose   domains run  from  $0$ to $\rho_0$ and from 
$\rho_0$ to $\infty$.  The dimensionless parameter $\rho_0>0$, is  chosen such that it satisfies simultaneously the conditions $\xi_\epsilon^{-1}\ll\rho_0\ll1$ and $(\eta_\epsilon/\xi_\epsilon^2)^{\nicefrac{1}{3}}\ll \rho_0$ 
with  $\eta_\epsilon=\varkappa_+k_-/m_\epsilon^2$.  In the former  integral  we  
Taylor expand the functions given in Eq.~(\ref{parameters}):  $X(\varphi)\approx-\rho^2[\psi^{\prime}(\varphi)]^2$ and  
$m_*^2(\varphi)\approx m_\epsilon^2\left[1+\frac{\xi_\epsilon^2\rho^2}{3}[\psi^{\prime}(\varphi)]^2\right]$. Afterward,  we perform the change of  variable  $s=\rho\xi_\epsilon$  and  extend  the  resulting integration 
limit $\rho_0\xi_\epsilon\to\infty$. No relevant contribution comes from the  integral defined in  $[\rho_0,\infty)$. Therefore, in the strong field regime $\xi_\epsilon\gg1$ [$\eta_\epsilon \ll \xi_\epsilon^2$], the function $\Delta(\infty)$ [see Eq.~(\ref{differneceformfactors})]  
is  well approximated  by
\begin{eqnarray}
\Delta(\infty)=-\alpha_\epsilon m_\epsilon^2 \int_{-\infty}^\infty d\varphi\int_{-1}^1dv \left[\frac{\mathrm{Gi}^\prime(x)}{x}+i\frac{\mathrm{Ai}^\prime(x)}{x}\right].
\label{strongfieldapproach}
\end{eqnarray} Here, $x=\left(6/[\vert\zeta_\epsilon(\varphi)\vert(1-v^2)]\right)^{\nicefrac{2}{3}}$,  $\mathrm{Gi}(x)$ and $\mathrm{Ai}(x)$ are the Scorer and Airy functions of first kind \cite{NIST},  respectively. 
In this context, $\zeta_\epsilon(\varphi)=3 \chi_\epsilon\psi^\prime(\varphi)/2$,  with   $\chi_\epsilon=\xi_\epsilon\eta_\epsilon$, refers to the  pulse-modulated 
nonlinear parameter associated  with the MCP vacuum.

We proceed our analysis by inserting the imaginary part of Eq.~(\ref{strongfieldapproach}) into Eq.~(\ref{rot}). As a consequence of the  relation   
$\mathrm{Ai}^\prime(z)=-\frac{z}{\pi\sqrt{3}}K_{\nicefrac{2}{3}}\left(\frac{2}{3}z^{{\nicefrac{3}{2}}}\right)$, with a  modified Bessel function  $K_{\nu}(z)$ \cite{NIST}, the following representation for 
the rotation angle is found
\begin{equation}\label{rotationstrong}
\begin{split}
&\vert\delta\vartheta(\epsilon,m_\epsilon)\vert\approx \frac{1}{2}\sin(2\vartheta_0)\frac{\alpha_\epsilon m_\epsilon^2}{\sqrt{3}\pi(\varkappa_+k_-)}\\ &\qquad\qquad\qquad\times\left \vert\int_{-\infty}^\infty d\varphi
\int_{0}^1dv\ K_{\nicefrac{2}{3}}\left(\frac{4}{\vert\zeta_\epsilon(\varphi)\vert}\frac{1}{1-v^2}\right)\right \vert.
\end{split}
\end{equation}Likewise, by substituting the real part of Eq.~(\ref{strongfieldapproach}) into Eq.~(\ref{elip}), we find for the ellipticity 
\begin{equation}\label{ellipstrong}
\begin{split}
&\vert\psi(\epsilon,m_\epsilon)\vert\approx \frac{1}{2}\sin(2\vartheta_0)\frac{\alpha_\epsilon m_\epsilon^2}{6^{\nicefrac{2}{3}}(\varkappa_+k_-)}\left \vert\int_{-\infty}^\infty d\varphi\ \vert\zeta_\epsilon(\varphi)\vert^{\nicefrac{2}{3}} \right.\\
&\qquad\qquad\times\left.\int_{0}^1dv (1-v^2)^{\nicefrac{2}{3}}\mathrm{Gi}^\prime\left[\left(\frac{6}{\vert\zeta_\epsilon(\varphi)\vert}\frac{1}{1-v^2}\right)^{\nicefrac{2}{3}}\right]\right \vert.
\end{split}
\end{equation} Eqs.~(\ref{rotationstrong}) and (\ref{ellipstrong}) are   used in the next section to  estimate the projected bounds in the parameter space of MCPs. Note that a numerical comparison 
between these expressions and the corresponding ones resulting from Eqs.~(\ref{rot}), (\ref{elip}) and  (\ref{differneceformfactors}) agrees within a few percent  
whenever $\xi_\epsilon\gg1$ and  $\zeta_\epsilon^{\nicefrac{1}{3}}\ll\xi_\epsilon$,  in agreement with the conditions imposed above  Eq.~(\ref{strongfieldapproach}).

In addition, further insights can be gained by  restricting $\zeta_\epsilon=3\chi_\epsilon/2$ to some asymptotic limits. We start with the case   $\zeta_\epsilon\ll1$. To be  consistent  
with $\xi_\epsilon\gg1$ the parameter $\eta_\epsilon$ must be restricted to $\eta_\epsilon\ll 2/(3\xi_\epsilon)$. In this limit  we can  exploit the  asymptotes $K_\nu(z)\sim \sqrt{\frac{\pi}{2 z}}e^{-z}$  
and  $\mathrm{Gi}(z)\sim\frac{1}{\pi z}$  \cite{NIST}. With these approximations, the  integrations  over $v$ can be performed in both observables. The expression for the ellipticity becomes particularly 
simple and can be computed exactly. Conversely, the calculation of the integral contained in the rotation angle requires  additional approximations.  
To this end, we first apply the change of variable  $w=\left(1-v^2\right)^{-1}$  and note that the region $w\sim1$ provides the essential contribution. This leads to 
\begin{eqnarray}\label{observablessmallchi1}
\begin{split}
&\vert\delta\vartheta(\epsilon,m_\epsilon)\vert\approx \frac{1}{2}\sin(2\vartheta_0)\frac{\alpha_\epsilon m_\epsilon^2}{8\sqrt{6}(\varkappa_+k_-)}\int_{-\infty}^\infty d\varphi
\  \vert\zeta_\epsilon(\varphi)\vert e^{-\frac{4}{\vert\zeta_\epsilon(\varphi)\vert}},\\
&\vert\psi(\epsilon,m_\epsilon)\vert\approx\frac{1}{2}\sin(2\vartheta_0)\frac{2\alpha_\epsilon m_\epsilon^2}{135 \pi(\varkappa_+k_-)}\int_{-\infty}^\infty d\varphi\ \vert \zeta_\epsilon(\varphi)\vert^2.
\end{split}
\end{eqnarray}The situation is  different when   $\xi_\epsilon\gg\zeta_\epsilon^{\nicefrac{1}{3}}\gg  1$.  
In this case,     $K_\nu(z)\sim\frac{\Gamma(\nu)}{2}\left(\frac{2}{z}\right)^\nu$  and  
$\mathrm{Gi}(z)\sim\frac{1}{ 2\pi\  3^{\nicefrac{2}{3}}}\Gamma\left(\frac{1}{3}\right)+\frac{1}{2\pi\ 3^{\nicefrac{1}{3}}}\Gamma\left(\frac{2}{3}\right) z$  applies  \cite{NIST}: 
\begin{eqnarray}\label{observableslargechi1}
\begin{split}
&\vert\delta\vartheta(\epsilon,m_\epsilon)\vert\approx\sqrt{3}\vert\psi(\epsilon,m_\epsilon)\vert,\\ 
&\vert\psi(\epsilon,m_\epsilon)\vert\approx \frac{1}{2}\sin(2\vartheta_0)\frac{2^{\nicefrac{1}{3}}\alpha_\epsilon m_\epsilon^2\Gamma^2(\frac{2}{3})}{7\sqrt{\pi}(\varkappa_+k_-)\Gamma(\frac{1}{6})} \left\vert\int_{-\infty}^\infty d\varphi
\  \vert  \zeta_\epsilon (\varphi)\vert^{\nicefrac{2}{3}}\right\vert,
\end{split}
\end{eqnarray}where $\Gamma(x)$ denotes the Gamma function. We remark that, if the external background  is  a  constant crossed field [$\psi^\prime(\varphi)=1$] which extends over $\Delta x_-$, 
the ellipticity  in  Eq.~(\ref{observablessmallchi1})  agrees with Eq.~(50) in Ref.~\cite{Dinu:2013gaa}, provided the distance traveled by the probe is given by  $d=\sqrt{2}\Delta x_-$ and  $\vartheta_0=\pi/4$.

So far,  no restriction has been imposed on the  field profile function $\psi^\prime(\varphi)$.  To proceed further,  we  take it  of the form
\begin{equation}
\psi^\prime(\varphi)=e^{-\frac{\varphi^2}{2\Delta\varphi^2}}\sin(\varphi).\label{fieldprofilefunction}
\end{equation}Here, $\Delta\varphi=\pi\mathpzc{N}\,/\sqrt{2\ln(2)}$ with $\mathpzc{N}$ referring to the number of oscillation cycles within the Gaussian envelop (FWHM).  
We insert this function  into the expression for the ellipticity  [see Eq.~(\ref{observablessmallchi1})] to  establish 
\begin{eqnarray}
\vert\psi(\epsilon,m_\epsilon)\vert\approx \frac{1}{2}\sin(2\vartheta_0)\frac{\alpha_\epsilon m_\epsilon^2 \zeta_\epsilon^2 \Delta\varphi}{135\sqrt{\pi}(\varkappa_+k_-)}\left[1-e^{-\Delta\varphi^2}\right].\label{psiexact}
\end{eqnarray} The expression given in  Eq.~(\ref{psiexact})  is valid if simultaneously  $\xi_\epsilon\gg1$ and $\zeta_\epsilon\ll1$. For $\xi_\epsilon=10$, $\zeta_\epsilon=0.15$ [$\zeta_\epsilon=3/2$] and  
$\Delta\varphi=4\pi$, it  differs from the exact formula Eq.~(\ref{elip})--with Eqs.~(\ref{differneceformfactors}) and (\ref{parameters}) included--by only  $0.2\%$ [$13\%$].

The integrals which remain  in  $\vert\delta\vartheta(\epsilon,m_\epsilon)\vert$ [see Eq.~(\ref{observablessmallchi1})] cannot be computed  analytically. To approximate them, we write
\begin{equation*}
\int_{-\infty}^\infty d\varphi\ldots=2\zeta_\epsilon\sum_{n=1}^{\infty}(-1)^{n-1}\int_{(n-1)\pi}^{n\pi} d\varphi\ \psi^\prime(\varphi)e^{\frac{4}{\zeta_\epsilon}\frac{(-1)^n}{\psi^\prime(\varphi)}},
\end{equation*} assume that $\zeta_\epsilon\ll1$, and  apply  the Laplace method. To  this end   we first note that the integrands 
vanish  at the boundaries and that the main contributions in the series arise from those values of  $\varphi$ which satisfy the condition $(n-1)\pi<\varphi< n\pi<\sqrt{2}\Delta\varphi$.  
Therefore,   the series  can be cut off  at  $N_{\mathrm{max}}=\lfloor1+\mathpzc{N}\,/\ln(2)\rfloor$, where $\lfloor x\rfloor$ refers to the  integer value of $x$.  
In addition, for the stationary points the condition  $\Delta\varphi^2/\varphi\gg1$ applies. Hence, we can use  the approximation   $\varphi\approx (2n-1)\pi/2$ with $n\in\mathbb{N}$. As a consequence,  
\begin{equation}\label{integralsmallchipara}
\int_{-\infty}^\infty d\varphi\ldots\approx2\zeta_\epsilon^{\nicefrac{3}{2}}\sqrt{\frac{\pi}{2}}\ \sum_{n=1}^{N_{\mathrm{max}}}\frac{1}{\gamma_n}e^{-\frac{4}{\zeta_\epsilon}\gamma_n},
\end{equation}with the parameter  $\gamma_n=\exp[(2n-1)^2\pi^2/(8\Delta\varphi^2)]$. We insert  this  approximation into $\vert\delta\vartheta(\epsilon,m_\epsilon)\vert$ [see Eq.~(\ref{observablessmallchi1})] 
and assume  $\mathpzc{N}\approx 5$.   Then, the main contribution arises  from the first term of the series above. Explicitly, 
\begin{equation}
\vert\delta\vartheta(\epsilon,m_\epsilon)\vert\approx \frac{1}{2}\sin(2\vartheta_0)\frac{1}{8}\frac{\alpha_\epsilon m_\epsilon^2\zeta_\epsilon^{\nicefrac{3}{2}}}{(\varkappa_+k_-)} \sqrt{\frac{\pi}{3}}
\frac{1}{\gamma_1}e^{-\frac{4}{\zeta_\epsilon}\gamma_1}.\label{rotsmalchi}
\end{equation}This  result   provides evidence that the photo-production probability of a pair of MCPs is  suppressed as  $\sim \exp(-4\gamma_1/\zeta_\epsilon)$, whenever 
$\xi_\epsilon\gg1$ and $\zeta_\epsilon\ll1$. This is   expected because  the damping factors of the probe $\kappa_{1,2}$ [see 
above Eq.~(\ref{difference})] represent the probability of producing a pair from  the respective propagating mode \cite{Meuren:2014uia}. 

The integration which remains in Eq.~(\ref{observableslargechi1}) can be estimated by replacing  the periodic term $\vert\sin(\varphi)\vert^{\nicefrac{2}{3}}$ by its average value,  
$\langle \vert\sin(\varphi)\vert^{\nicefrac{2}{3}} \rangle=3\sqrt{\frac{3}{\pi}}\frac{\Gamma(\frac{2}{3})}{\Gamma(\frac{1}{6})}$. Correspondingly, the ellipticity [rotation angle] acquires the form
\begin{equation}\label{ellipbigchi}
\begin{split}
&\vert\psi(\epsilon,m_\epsilon)\vert\approx \frac{1}{2}\sin(2\vartheta_0)\frac{18}{7\sqrt{\pi}}\frac{\alpha_\epsilon m_\epsilon^2 }{(\varkappa_+k_-)}\frac{\Gamma^3(\frac{2}{3})}{\Gamma^2(\frac{1}{6})}\left(\frac{\zeta_\epsilon}{2}\right)^{\nicefrac{2}{3}}\Delta\varphi,\\
&\vert\delta\vartheta(\epsilon,m_\epsilon)\vert\approx\sqrt{3}\vert\psi(\epsilon,m_\epsilon)\vert.
\end{split}
\end{equation}These analytical results were derived by assuming that $\xi_\epsilon \gg \zeta_\epsilon^{1/3} \gg 1$. The expression for the ellipticity
[rotation angle] given in Eq.~(\ref{ellipbigchi})  agrees  with Eq.~(\ref{ellipstrong}) [Eq.~(\ref{rotationstrong})] within an accuracy of $<19\%$ [$<3\%$] 
 if $\zeta_\epsilon>10^3$  for $\Delta\varphi = 4\pi$.  
  
Some comments   are in order.  First of all, while  Eq.~(\ref{psiexact}) is exact  with respect to the integration over $\varphi$, the approximations used to obtain  Eqs.~(\ref{rotsmalchi}) and (\ref{ellipbigchi}) prevent us 
from taking the monochromatic  limit [$\Delta\varphi\to\infty$]  directly. Instead, this limiting  case can be derived by  noting that the 
integrands  in  $\vert\delta\vartheta(\epsilon,m_\epsilon)\vert$ [see Eq.~(\ref{observablessmallchi1}) and   Eq.~(\ref{observableslargechi1})] are $\pi$-periodic. In  this situation, we have  $\int_{-\infty}^\infty d\varphi\ldots=2N\int_0^\pi d\varphi\ldots$ with $N\to\infty$ and  thus,
\begin{equation*}
\int_{-\infty}^\infty d\varphi\ldots=2\pi N \left\{\begin{array}{cc}\displaystyle
\frac{1}{\sqrt{2\pi}}\zeta_\epsilon^{\nicefrac{3}{2}}e^{-\frac{4}{\zeta_\epsilon}}&  \zeta_\epsilon\ll1\\ \\ \displaystyle
\left\langle\vert\sin(\varphi)\vert^{\nicefrac{2}{3}}\right\rangle &  \zeta_\epsilon\gg1,
\end{array}\right.
\end{equation*}where the result  for $\zeta_\epsilon\ll1$ has been quoted from Ref.~\cite{Meuren:2014uia}. Hence, we only need to carry out the respective replacements  $\exp(-4\gamma_1/\zeta_\epsilon)/\gamma_1\to Ne^{-\frac{4}{\zeta_\epsilon}}$ and $\Delta\varphi\to 2N\sqrt{\pi/3}$  in 
Eqs.~(\ref{rotsmalchi}) and (\ref{ellipbigchi}),  to  establish  the asymptotic behaviors of $\vert\delta\vartheta(\epsilon,m_\epsilon)\vert$ and $\vert\psi(\epsilon,m_\epsilon)\vert$  in the 
monochromatic limit.

\subsection{Leading behavior at weak fields  $\xi_\epsilon\ll1$ \label{smallxi}}

In the  regime $\xi_\epsilon\ll 1$, the pulse [see Eq.~(\ref{extfield}) with $\psi_2(\varphi)=0$] constitutes  a small perturbation. The leading order contribution of the corresponding  
expansion $\sim\xi^2_\epsilon$  in the  polarization tensor $\Pi_{\mu\nu}(x,x^\prime)$ describes the scattering of a probe photon by a  photon of  the laser pulse [photon-photon scattering]. Since 
the light-by-light scattering cross section is maximized in the vicinity of  the pair creation threshold [$n_*=2m_\epsilon^2/\vert\varkappa_+k_-\vert\approx1$],  we can anticipate a strong dispersive effect  
around  the  threshold mass for MCPs $m_1\equiv\sqrt{\frac{1}{2}\vert\varkappa_+k_-\vert}$. This is understandable because, for such  energies [$\omega_{\pmb{k}}\approx m_\epsilon^2/\varkappa_0\pm\delta \omega$ 
with $m_\epsilon^2/\varkappa_0\gg \delta\omega>0$], the probe photons coexist  with quasi-resonant fluctuations of the $\mathpzc{q}_\epsilon^+\mathpzc{q}_\epsilon^-$  field.  In contrast, far from the threshold  
[$n_*\to \infty$ and $n_*\to0$],   dispersive effects are predicted to be much less  pronounced. Accordingly, we can expect less stringent bounds for  masses  far away from the threshold mass.

Above the pair production threshold $1>n_*$ the imaginary part of the polarization operator is different from zero and the vacuum
becomes dichroic.  Below  threshold,  absorptive phenomena may also occur,  but such  processes are less likely since they  are  linked to higher order Feynman diagrams involving--at least--two photons of 
the external pulse.  Contributions of higher order processes $k+n\varkappa\to \mathpzc{q}_\epsilon^++ \mathpzc{q}_\epsilon^-$ with  $n>1$ are beyond the scope of this work [see Refs.~\cite{Villalba-Chavez:2014nya,Villalba-Chavez:2015xna} 
for more details].

Let us now  specialize the observables  [see Eqs.~(\ref{rot}) and (\ref{elip})] to the case  $\xi_\epsilon \ll 1$. As before,  we  apply the  change of variable $\tau=4\rho/[\vert\varkappa_+k_-\vert(1-v^2)]$. 
The resulting  dressing  factor in the effective mass $m_*^2-m_\epsilon^2\sim \xi_\epsilon^2$ [see Eq.~(\ref{parameters})] becomes very small in comparison with the leading order term $m_\epsilon^2$, allowing us 
to make an expansion  in  $\xi_\epsilon^{2}$ which turns out to be valid whenever $n_*\ll\xi_\epsilon^{-2}$. Afterward,  the variable $\varphi$ is  integrated  out  using the pulse profile function [see 
Eq.~(\ref{fieldprofilefunction})]. Correspondingly,
\begin{eqnarray}
\Delta(\infty)=\frac{\alpha_\epsilon}{\pi}m_\epsilon^2 \xi_{\epsilon}^2 \int_{-1}^1dv\int_0^{\infty} \frac{d\rho}{\rho}\int_{-\infty}^\infty d\varphi X(\varphi) \exp\left[-\frac{2in_{*}\rho}{1-v^2}\right],
\label{differneceformfactorsllscattering}
\end{eqnarray}where 
\begin{equation}
\begin{split}
&\int_{-\infty}^\infty d\varphi X(\varphi)=2\sqrt{\pi}\rho^2 \Delta\varphi\int_{0}^1 dy\int_0^1 dy^\prime e^{-\frac{\rho^2 (y-y^\prime)^2}{\Delta\varphi^2}}\\
&\qquad\qquad\times(y^\prime-1)y\left\{\cos(2\rho\sigma[y-y^\prime])-\exp\left(-\Delta\varphi^2\right)\right\}.
\end{split}\label{integrationvarphi1}
\end{equation}Here, we introduced the parameter $\sigma=\varkappa_+k_-/\vert\varkappa_+k_-\vert$.  Three out of the four integrations can be carried out analytically. 
To this end, we first  introduce  two new variables $s^{-1}=y-y^\prime$ and $z=y+y^\prime$ and carry out the  integrations over $z$ and $\rho$. With help of the  shorthand notation  
$\ell_s=n_{*}s/[\sigma (1-v^2)]$, we find a two-fold integral representation for the real and the imaginary part [see   Eq.~(\ref{differneceformfactorsllscattering})]
\begin{equation}\label{IMRE1}
\begin{split}
&\mathrm{Im}\Delta(\infty)=\frac{1}{4}\alpha_\epsilon (\varkappa_+k_-) \xi_{\epsilon}^2\Delta\varphi^2 \int_{0}^1dv(1-v^2)\int_1^\infty \frac{ds}{s^4}\\&\qquad\qquad\times\left\{e^{-\Delta\varphi^2\left(1+\ell_s\right)^2}+e^{-\Delta\varphi^2\left(1-\ell_s\right)^2}-2 e^{-\Delta\varphi^2\left(1+ \ell_s^2\right)}\right\},\\
\end{split}
\end{equation}
\begin{equation}\label{IMRE2}
\begin{split}
&\mathrm{Re}\Delta(\infty)=\frac{1}{2\sqrt{\pi}} \alpha_\epsilon(\varkappa_+k_-)\xi_{\epsilon}^2\Delta\varphi^2 \int_{0}^1dv(1-v^2)\int_1^\infty \frac{ds}{s^4}\\&\qquad\times\left\{\frac{}{}\mathpzc{D}_F\left(\Delta\varphi\left[1+\ell_s\right]\right)-\mathrm{sig}(1-\ell_s)\mathpzc{D}_F\left(\Delta\varphi\left\vert1-\ell_s\right\vert\right)\right.\\
&\qquad\qquad-\left.2 e^{-\Delta\varphi^2}\mathpzc{D}_F(\Delta\varphi\ell_s)\ \right\},
\end{split}
\end{equation}where   $\mathpzc{D}_F(x)=e^{-x^2}\int_0^x dt e^{t^2}$ is the Dawson  function  \cite{NIST}. Now,  we  perform  in  
Eqs.~(\ref{IMRE1}) and (\ref{IMRE2}) the  changes of variables $x_1=\Delta\varphi\left[1+\ell_s\right]$, $x_2=\Delta\varphi\left[1-\ell_s\right]$  and $x_3=\Delta\varphi\ell_s$ in the first, second and third 
contribution, respectively. After an integration by parts with respect to  $v$,  the integral over $s$ is eliminated and we end up with the following expression for the rotation  angle
\begin{equation}\label{IMREFINAL1}
\begin{split}
&\vert\delta\vartheta(\epsilon,m_\epsilon)\vert=\frac{1}{4}\sin(2\vartheta_0)\alpha_\epsilon\xi_{\epsilon}^2\Delta\varphi^2\left\vert \int_{0}^1dv\ v (1-v^2)\right.\\&\qquad\times\left.\left[\frac{1-v^2}{2}\ln\left(\frac{1+v}{1-v}\right)+v\right]e^{-\Delta\varphi^2(1+\ell_1^2)}\sinh^2\left(\Delta\varphi^2\ell_1\right)\right\vert
\end{split}
\end{equation}and the induced ellipticity
\begin{equation}\label{IMREFINAL2}
\begin{split}
&\vert\psi(\epsilon,m_\epsilon)\vert=\frac{1}{2}\sin(2\vartheta_0)\frac{1}{4\sqrt{\pi}} \alpha_\epsilon\xi_{\epsilon}^2\Delta\varphi^2\left\vert \int_{0}^1dv\ v (1-v^2)\right.\\&\qquad\times \left[\frac{1-v^2}{2}\ln\left(\frac{1+v}{1-v}\right)+v\right]\left\{\frac{}{}\mathpzc{D}_F\left(\Delta\varphi\left[1+\ell_1\right]\right)\right.\\&\qquad\qquad\quad-\left.\left.\mathpzc{D}_F\left(\Delta\varphi\left[1-\ell_1\right]\right)-2 e^{-\Delta\varphi^2}\mathpzc{D}_F(\Delta\varphi\ell_1)\frac{}{} \right\}\right\vert.
\end{split}
\end{equation} The expressions  in Eqs.~(\ref{IMREFINAL1}) and (\ref{IMREFINAL2}) hold for the pulse shape given in Eq.~(\ref{fieldprofilefunction}) and 
apply whenever   $\xi_\epsilon\ll1$  and  $n_*\ll\xi_\epsilon^{-2}$. The numerical values provided by both expressions  agree with the exact results calculated 
from Eqs.~(\ref{rot}) and (\ref{elip}),  including  Eqs.~(\ref{differneceformfactors}) and (\ref{parameters}), within a few percent. 

It is interesting to deal with some special cases. Let us consider first  the rotation angle [see Eq.~(\ref{IMREFINAL1})]. Assuming 
the condition $\Delta\varphi^2>\Delta\varphi^2 n_*\gg1$,  one can use the approximation  $\sinh^2(\Delta\varphi^2\ell_1 )\approx\frac{1}{4}\exp[2\Delta\varphi^2\ell_1]$ and apply the Laplace 
method. Finally, Eq.~(\ref{rot}) leads to  the expression
\begin{eqnarray}
\begin{split}
&\vert\delta\vartheta(\epsilon,m_\epsilon)\vert\approx\frac{1}{4}\sin(2\vartheta_0)\frac{1}{8}\alpha_\epsilon\xi_{\epsilon}^2\Delta\varphi\sqrt{\pi} (1-\mathpzc{v}_1^2)^2\left\vert\left[\frac{1-\mathpzc{v}_1^2}{2}\right.\right.\\&\qquad\qquad\qquad\quad\times\left.\left.\ln\left(\frac{1+\mathpzc{v}_1}{1-\mathpzc{v}_1}\right)+\mathpzc{v}_1\right]\left\{\frac{1}{2}+\frac{1}{2}\mathrm{Erf}\left(\Delta\varphi\mathpzc{v}_1^2\right)\right\}\right\vert,
\end{split}\label{Imdeltafinal}
\end{eqnarray} with   $\mathrm{Erf}(x)=\frac{2}{\sqrt{\pi}}\int_0^x dt \exp[-t^2]$ denoting the error function \cite{NIST}. This formula  applies as long as  the condition $\Delta\varphi^{-2}\ll n_*<1$  is satisfied. 
We point out that the quantity $\mathpzc{v}_1=(1-n_*)^{\nicefrac{1}{2}}$ defines  the relative speed of  the  final particle 
states in the center--of--mass frame.  In the  monochromatic limit [$\Delta\varphi\to\infty$],   the expression in Eq.~(\ref{Imdeltafinal}) contained 
within the curly brackets reduces to  the unit step  function  $\Theta(\mathpzc{v}_1^2)$. We note that, for the test  parameters  $\xi_\epsilon=0.1$, $\mathpzc{n}_*=0.02$ and  
$\Delta\varphi=4\pi$, the relative difference  between this expression and the exact formula Eq.~(\ref{rot})--with Eqs.~(\ref{differneceformfactors}) 
and (\ref{parameters}) included--is smaller than $3\%$.

As $\Delta\varphi^2 n_*\ll1<\Delta\varphi^2$ implies  $\sinh(\Delta\varphi^2\ell_1)\approx\Delta\varphi^2\ell_1$  [see  Eq.~(\ref{IMREFINAL1})], we find that
 $\vert\delta\vartheta(\epsilon,m_\epsilon)\vert\sim n_*^2\Delta\varphi^6\exp(-\Delta\varphi^2)$  is  exponentially suppressed, which  indicates that in this regime 
 vacuum  dichroism  tends to vanish. 
 
We point out that Eq.~(\ref{psiexact})  also applies if $\xi_\epsilon\ll1$ and $1\ll\Delta\varphi n_*$. To show this, we use 
$\mathpzc{D}_F(\Delta\varphi(1\pm \ell_1))\approx\pm \mathpzc{D}_F(\Delta\varphi\ell_1)\approx\pm1/(2\Delta\varphi\ell_1)$,
implying $\int_0^1dv\ldots\approx\frac{4}{15}(1-e^{-\Delta\varphi^2})$  in Eq.~(\ref{IMREFINAL2}). In the regime   $\Delta\varphi n_*\ll1$  
we apply the change of variable $t=1-v^2$ and  introduce  a splitting parameter $t_0$ with 
$\Delta\varphi n_* \ll t_0\ll1$.  Afterward, the $t$ integration is divided into ranges  from $0$ to $t_0$ and from $t_0$ to $1$. In the first region, we have $t\ll1$ and a Taylor expansion is feasible.  
After an integration by parts, we  obtain
\begin{equation}
\int_0^{t_0}dt\ldots\approx t_0 \Delta\varphi n_*\left\{\mathpzc{D}_F^\prime(\Delta\varphi)-e^{-\Delta\varphi^2}\right\}.\label{a1}
\end{equation}Since in  the second range  $\Delta\varphi n_*\ll t$, we can expand the  expression contained in the curly brackets [see  $\vert\psi(\epsilon,m_\epsilon)\vert$ in  Eq.~(\ref{IMREFINAL2})] in $\Delta\varphi n_*/t$. 
Hence,
\begin{equation}\label{a2}
\begin{split}
&\int_{t_0}^1dt\ldots\approx 2\Delta\varphi n_*\left\{\mathpzc{D}_F^\prime(\Delta\varphi)-e^{-\Delta\varphi^2}\right\}\\ &\qquad\qquad\qquad\times\int_{t_0}^1dt\ t \left[\frac{\sqrt{1-t}}{t}+\frac{1}{2}\ln\left(\frac{1+\sqrt{1-t}}{1-\sqrt{1-t}}\right)\right]
\end{split}
\end{equation} To leading order, the remaining integral reads  $\int_{t_0}^1\ldots\approx(1-t_0/2)$. After combining both parts [see Eqs.~(\ref{a1}) and (\ref{a2})], 
the ellipticity becomes
\begin{equation}
\begin{split}
&\vert\psi(\epsilon,m_\epsilon)\vert\approx \frac{1}{2}\sin(2\vartheta_0)\frac{\alpha_\epsilon\xi_\epsilon^2n_*\Delta\varphi^3 }{4\sqrt{\pi}}\\ &\qquad\qquad\qquad\qquad\times \left\vert1-2\Delta\varphi\mathpzc{D}_F(\Delta\varphi)-e^{-\Delta\varphi^2}\right\vert,
\end{split}\label{lastexpression}
\end{equation}where $\mathpzc{D}_F^\prime(\Delta\varphi)=1-2\Delta\varphi\mathpzc{D}_F(\Delta\varphi)$ has been used \cite{NIST}. The monochromatic limit [$\Delta\varphi\to\infty$] 
 can be   investigated through  $\mathpzc{D}_F(\Delta\varphi)\approx\frac{1}{2\Delta\varphi}-\frac{1}{4\Delta\varphi^3}$, in which case the induced ellipticity reads 
 $\vert\psi(\epsilon,m_\epsilon)\vert\approx\frac{1}{2}\sin(2\vartheta_0)\frac{1}{8\sqrt{\pi}}\alpha_\epsilon\xi_\epsilon^2\Delta\varphi n_*$. Finally, as a check, we found that 
 for   $\xi_\epsilon=0.1$, $\mathpzc{n}_*=0.02$ and  $\Delta\varphi=4\pi$, the outcomes from   Eq.~(\ref{lastexpression}) and the exact formula Eq.~(\ref{elip})--with 
 Eqs.~(\ref{differneceformfactors}) and (\ref{parameters}) included--agree within an accuracy of $0.1\%$.

\section{Experimental prospects}

We start by analyzing the HIBEF experiment proposed in  \cite{Schlenvoigt}, which is based on a Petawatt laser with $\varkappa_0\approx1.55\ \rm eV$ [$\lambda_0=800\ \rm  nm$], a repetition rate  of $1\ \rm Hz$, a temporal pulse length of about $30 \ \rm fs$ [$\Delta\varphi\approx11\pi$], 
and   a peak intensity $I\approx 2\times10^{22}\ \rm  W/cm^2$  corresponding to  $\xi\approx 69$. The probe beam  will be produced by the European  x-ray free electron laser [$\omega_{\pmb{k}}=12.9\ \mathrm{keV}$,  
$\mathcal{N}_{\mathrm{in}}\approx 5\times 10^{12}$ photons per shot], the  transmission coefficient of the optics is  $\mathcal{T}=0.0365$. In this experiment [$\vartheta_0=\pi/4$] an ellipticity 
$\vert\psi_{\mathrm{QED}}\vert=(9.8\pm6.7)\times10^{-7}\ \rm rad$ would be detectable  \cite{Schlenvoigt}. Using   Eq.~(\ref{ellipstrong}), we infer that MCPs with relative coupling constant  
$\epsilon<1.3\times 10^{-3}$ would not be ruled out whenever $m_\epsilon\lesssim100\ \rm eV$. We  have arrived at this limit  by  assuming that the induced ellipticity due to MCPs  does not overpass  
the upper bound set by the QED signal.

\begin{figure*}[t]
\centering
\includegraphics[width=0.49\linewidth]{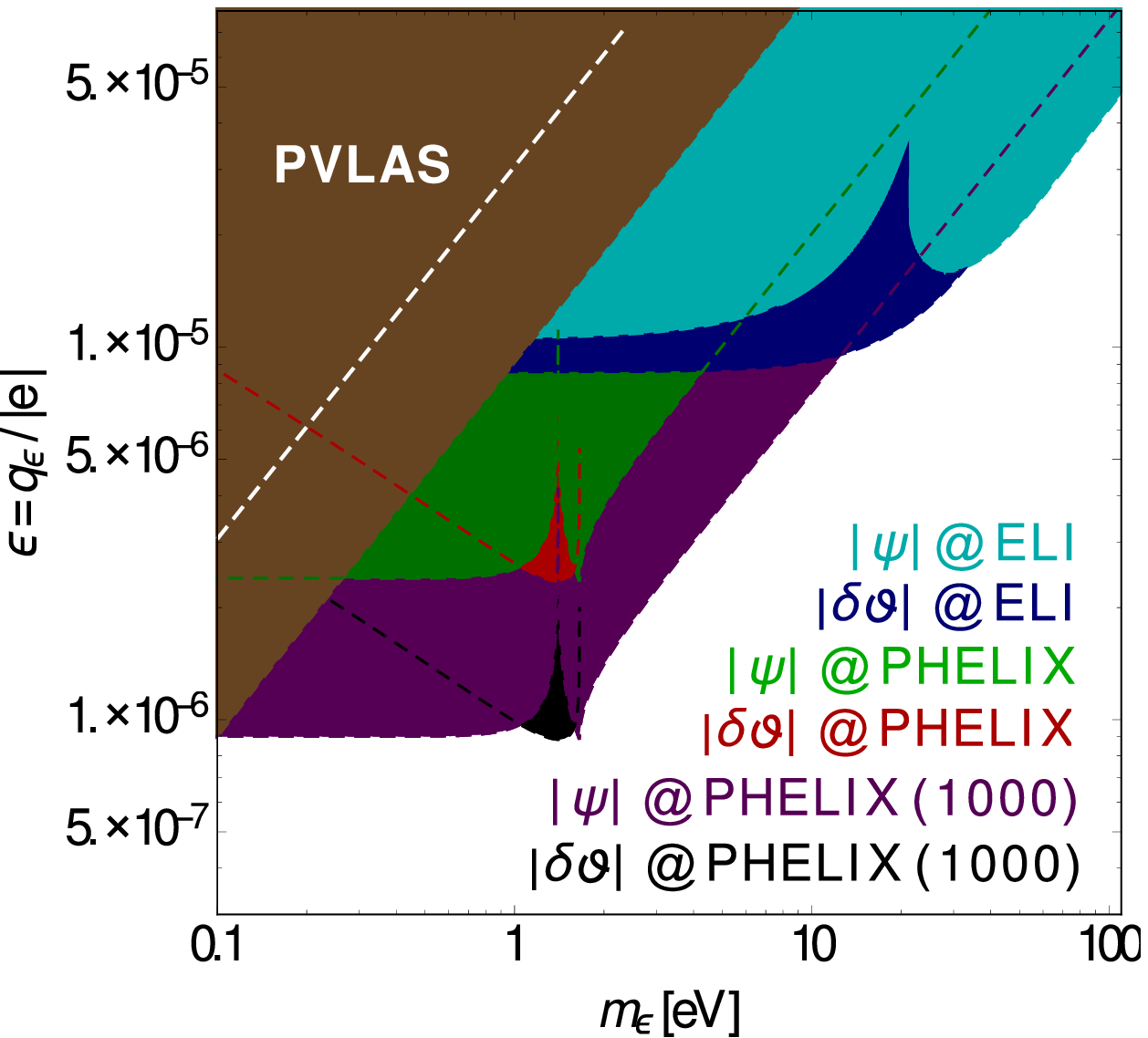}
\hfill
\includegraphics[width=0.49\linewidth]{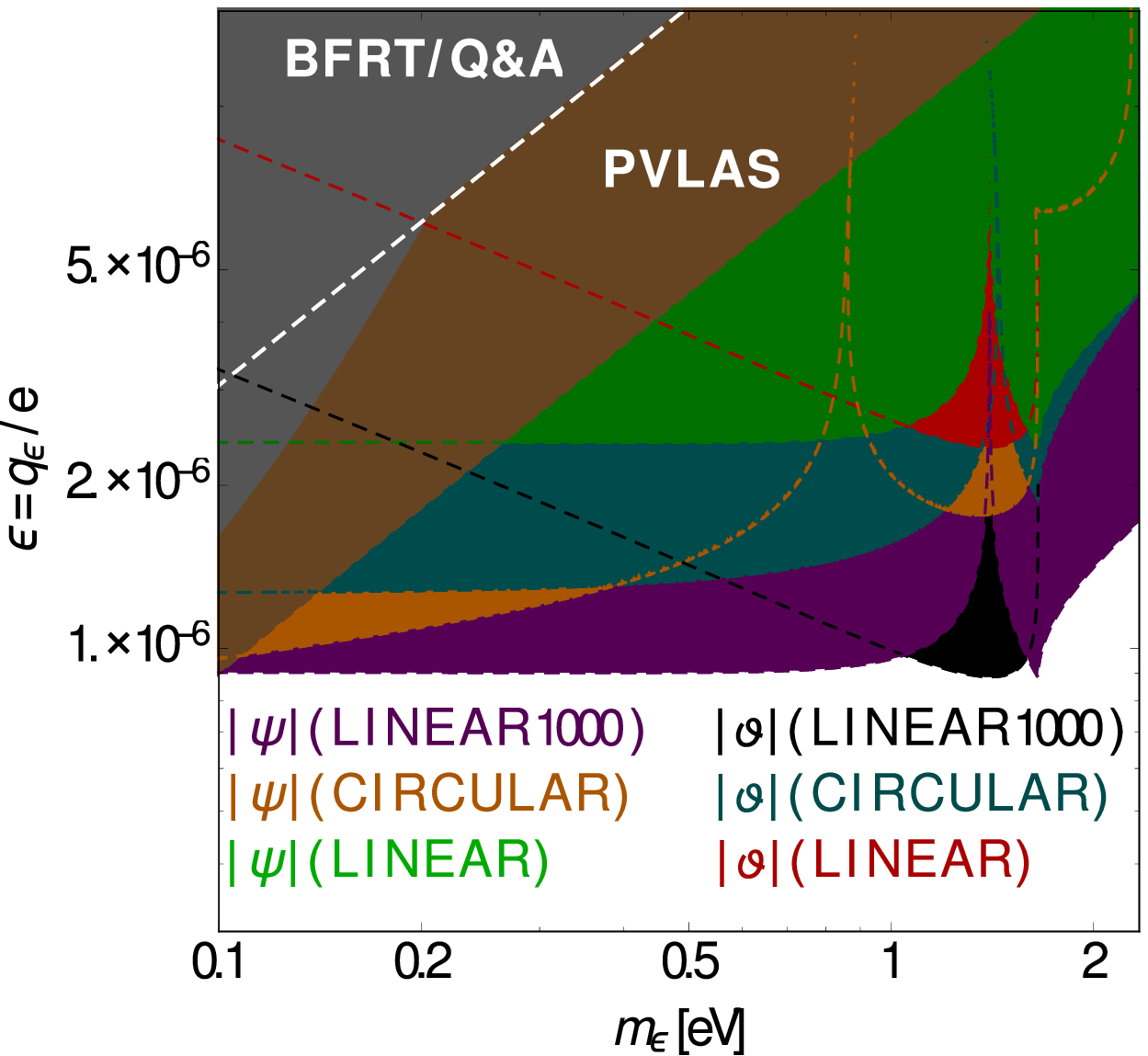}
\caption{\label{figure} Estimates of  constraints for MCPs  of  mass $m_\epsilon$ and relative coupling constant $\epsilon$ derived  from the absence of signals in  
a plausible polarimetric setup  assisted by a linearly polarized Gaussian laser pulse. In both panels, the white  dashed line  correspond to the expression 
$\xi_\epsilon=1$ which is evaluated with the  PHELIX parameters. The colored  regions in brown and gray are  exclusion areas  stemming  from various  
experimental  collaborations searching for rotation and ellipticity in constant magnetic fields  such as BFRT \cite{Cameron:1993mr},  PVLAS \cite{DellaValle:2013xs}  
and Q $\&$ A \cite{Chen:2006cd}.  The respective $95\%$  confidence  levels needed to recreate the BFRT and Q$\&$A results are summarized  in Ref.~\cite{Ahlers:2007qf}. }
\end{figure*}

As discussed below Eq.~(\ref{fouriertransf}), the energy scale $1/w_0$ associated with the waist size of the pulse $w_0$ limits the applicability of our method to the regime $m_\epsilon\gg w_0^{-1}$ 
[$w_0\approx 2\lambda_0\approx (0.12\ \rm eV)^{-1}$ for HIBEF]. For the detection of QED birefringence a 
detailed analysis of focussing effects has recently been carried out in Ref.~\cite{Karbstein:2015xra} based on an expression for the polarization operator which was obtained from the Euler-Heisenberg 
Lagrangian [see also \cite{Heinzl,DiPiazza:2006pr}]. It was shown there that focussing effects could notably improve the signal-to-noise ratio if probe photons which are scattered slightly away 
from the forward direction are analyzed.  Certainly, this  fact  might  be  beneficial in the search of MCPs as well.  However, we point out  that such a study would require to incorporate  transverse 
focusing effects in the polarization tensor. This  computation is challenging in the energy regimes considered here. Conversely, at low energies $\omega_{\pmb{k}}\varkappa_0\ll m_\epsilon^2$, the 
Euler-Heisenberg Lagrangian could be used, but this calculation is  beyond the scope of this work.

Next, let us  estimate  the projected limits resulting from a  technically feasible experiment in which the  rotation of the polarization plane [see Eq.~(\ref{rot})] and  the ellipticity [see Eq.~(\ref{elip})] 
are probed with an optical laser beam,   but none of them is  detected. In practice, the absence of these signals  provides certain upper limits  $\psi_{\mathrm{CL}\%}$,  $\delta\vartheta_{\mathrm{CL}\%}$
which are   understood  within  certain confidence levels, frequently corresponding to  $2\sigma$. Hereafter, we take $\psi_{\mathrm{CL}\%},\ \delta\vartheta_{\mathrm{CL}\%}\sim10^{-10}\ \mathrm{rad}$. This  choice is 
in  agreement with the experimental accuracies  with which both observables can nowadays be  measured in  the optical regime. Here, the projected sensitivities result from the  
inequalities $10^{-10}\ \mathrm{rad}>\vert\psi(\epsilon,m_\epsilon)\vert\quad$ and $10^{-10}\ \mathrm{rad}>\vert\delta\vartheta(\epsilon,m_\epsilon)\vert$. Firstly, we consider the nanosecond front-end of the PHELIX laser 
\cite{PHELIX}, [$\tau\approx20 \ \rm ns$, $\varkappa_0\approx1.17\ \rm eV$ implying  $\Delta\varphi\approx 5\times 10^6 \pi$,  $I_{\mathrm{max}}\approx 10^{16}\  \rm W/cm^2$, $\xi\approx 6.4\times 10^{-2}$, 
$w_0\approx 100-150 \ \mu \mathrm{m}$] combined with a frequency doubled probe beam [$\omega_{\pmb{k}}=2\varkappa_0=2.34\ \rm eV$], having a waist size and an intensity much smaller than
the corresponding ones of the strong laser field.

The  projected exclusion regions associated with this laser setup are shaded  in Fig.~\ref{figure}  in green  and red. These  should be trustworthy as long as the limits  lie much below   the  curve   
corresponding to $\xi_\epsilon=\epsilon m \xi/m_\epsilon=1$, i.e.  the white dashed line in the upper left corner. We remark that our potential exclusion bounds are  valid whenever the condition $m_\epsilon\gg w_0^{-1}$ is 
satisfied. This  translates into  $m_\epsilon\gg 1.3 \ \rm meV$. In line with this last aspect, we note that the pulse length  associated with PHELIX  is much larger than its  wave period  [$\tau\gg \varkappa_0^{-1}$]  and, 
furthermore,  satisfies the condition $w_0\gg\lambda_0$. Therefore, the electromagnetic field produced by this laser system can be treated theoretically as a monochromatic plane wave. It is also worth observing that 
the square of the intensity  parameter  associated with the  PHELIX beam  is  much smaller than unity $\xi^2\ll1$ [$\xi_\epsilon^2\ll1$ in the relevant parameter space]. Under these circumstances,  the observables  [see Eqs.~(\ref{rot}) and (\ref{elip})] are dominated by a  
dependence of the form  $\propto \xi^2 \Delta\varphi$,  as  can be read off from Eqs.~(\ref{IMREFINAL1}) and (\ref{IMREFINAL2}). This fact indicates  that--for $\omega_{\pmb{k}}\sim 1\ \rm eV$--large  sensitivities can be 
achieved  provided $\Delta\varphi$  compensates for the relative smallness of $\xi$. As we anticipated  in  Sec.~\ref{smallxi}, this enhancement is  particularly large in the vicinity of the threshold mass $m_1\approx1.64\ \rm eV$ because 
the cross section for photon-photon scattering is maximized nearby the pair creation threshold. Here, the projected bound coming from a  search  of the induced ellipticity turns 
out to be  $\epsilon< 2.8 \times 10^{-6}$. 

We note that the exclusion plot exhibits a discontinuity at the threshold mass [see discussion below Eq.~(\ref{Imdeltafinal})]. Upper bounds for large masses can be derived  when higher order proce\-sses--such as 
the three photon reaction--are taken into account \cite{Villalba-Chavez:2014nya,Villalba-Chavez:2015xna}. The effects resulting from this phenomenon are summarized  in the right panel of 
Fig.~\ref{figure} [orange area]. This outcome as well as the one in darker cyan for the rotation angle  were obtained previously by  assuming the strong field as a circularly polarized wave 
and considering a procedure beyond the Born approximation \cite{Villalba-Chavez:2014nya,Villalba-Chavez:2015xna}.  We note that in the case of circular polarization a slightly more stringent 
bound  of $\epsilon< 1.9 \times 10^{-6}$ at $m_1\approx1.64\ \rm eV$ results from the induced ellipticity. 

Both panels include  regions colored in purple and  black  labeled by PHELIX1000. These excluded areas have been determined  by using the PHELIX parameters given above but supposing that the signals  
gain sensitivity by a factor of  $\sim50$. This could be achieved if a series of plasma mirrors induces $1000$ crossings of the two beams as suggested by Tommasini et al. \cite{Tommasini:2009nh}. This 
method is feasible for intensities below $\sim10^{19}\ \rm W/cm^2$ and would  require a collision angle very close  to $\pi$. Besides, the  mirrors should exceed   
the waist size of the  pulse in order to avoid diffractive  distortions; for further details see \cite{Tommasini:2009nh}. Using the same sensitivity of  $\sim 10^{-10}$ as above, the exclusion limit 
is pushed down to $\epsilon<8.8\times 10^{-7}$  at the threshold mass  $m_1\approx1.64\ \rm eV$  [for all projected sensitivities we assume a counter propagating geometry $\varkappa_+k_-=2\varkappa_0\omega_{\pmb{k}}$ 
and an initial polarization angle  $\vartheta_0=\pi/4$].

As a last scenario, we consider the envisaged parameters at ELI:  $\tau\approx 13 \ \rm fs$, $\varkappa_0\approx1.55\ \rm eV$ 
[$\lambda_0=800\ \rm  nm$] corresponding to    $\Delta\varphi\approx4\pi$,  $I\approx 10^{25}\ \rm W/cm^2$, $\xi\approx 1.5\times 10^{3}$. Here, we analyze  the results   
taking the probe beam with  doubled frequency $\omega_{\pmb{k}}=2\varkappa_0=3.1\ \rm eV$, a waist size  and an  intensity  much smaller than the one of  the strong laser field, 
whereas $\psi_{\mathrm{CL}\%},\ \delta\vartheta_{\mathrm{CL}\%}\sim10^{-10}\ \mathrm{rad}$. Furthermore, a single-crossing geometry is assumed again.
The projected exclusion areas  are  shaded  in the left panel of  Fig.~\ref{figure} in cyan and blue. Since the field of the pulse  at ELI is expected to be strongly focused [$w_0\sim \lambda_0$], 
the  estimates associated with this setup  are expected to be reasonable as long as $m_\epsilon\gg 0.1 \ \rm eV$ and the upper limit of $\epsilon$ lies  much above  the  curves   corresponding 
to $\xi_\epsilon=1$ and $\zeta_{\epsilon}^{\nicefrac{1}{3}}=\xi_\epsilon$. [Note that these curves  lie  far below  the region  encompassed by the figure.]   We observe  that,  in the ELI  scenario, the path of the projected exclusion bounds resembles those 
established from  experiments driven  by constant magnetic fields \cite{Gies:2006ca,Ahlers:2007rd,Ahlers:2007qf}.

\section{Conclusions}

We have studied the prospects that laser-based experiments, designed to detect vacuum birefringence,  offer for  probing  hypothetical degrees of freedom with a tiny fraction of  
the electron charge. Throughout this investigation, we have indicated that  the vacuum of  MCPs might induce ellipticity and rotation on the incoming polarization plane,  even 
though the probe photon energy is much below the  threshold of electron-positron pair production. In such a scenario,  the  transmission probability through an analyzer set 
crossed to the  initial polarization direction  would not be  determined solely  by the QED ellipticity but also by the ellipticity and the rotation angle   induced by MCPs.  We 
have argued   that  a slightly modified version of the proposed polarimeter for a x-ray probe would  allow for measuring  both  signals separately. The projected bounds 
resulting from this analysis will depend on the choice of the wave profile.  In contrast to previous studies,  the treatment presented here  has  taken  into account the effects resulting 
from a  Gaussian envelop.  With the help of contemporary techniques based on  plasma mirrors, polarimetric studies driven by an optical laser pulse of moderate intensity 
[$\sim 10^{16}\ \rm W/cm^2$] might allow for  excluding  MCPs with $\epsilon>9\times 10^{-7}$  and  masses $0.1\ \mathrm{eV} \leqslant m_\epsilon <1.5\  \mathrm{eV}$, a region which 
has not been discarded so far by experiments driven by constant magnetic fields and where the best model-independent cosmological limits--resulting from CMB data--are of the same 
order of magnitude \cite{Melchiorri}.

\vspace{0.006 in}
\begin{flushleft}
\textbf{Acknowledgments}
\end{flushleft}
\vspace{0.005 in}
The authors thank A. Di Piazza and S. Bragin for useful comments to  the manuscript. S. Villalba-Ch\'avez and C. M\"{u}ller gratefully acknowledge the funding by the German Research Foundation (DFG) under Grant No. MU 3149/2-1.

\end{document}